\documentclass[journal=jacsat,manuscript=article]{achemso}

\usepackage[version=3]{mhchem} % Formula subscripts using \ce{}
\usepackage{appendix}
\author{Carmine Anzivino}
\affiliation[University of Milan]
{Department of Physics ``A. Pontremoli", University  of Milan, via Celoria 16, 20133 Milan, Italy}
\author{Klejdis Xhani}
\affiliation[University of Milan]
{Department of Physics ``A. Pontremoli", University of Milan, via Celoria 16, 20133 Milan, Italy}
\author{Marina Carpineti}
\affiliation[University of Milan]
{Department of Physics ``A. Pontremoli", University of Milan, via Celoria 16, 20133 Milan, Italy}
\author{Stefano Verrastro}
\affiliation[University of Milan]
{Department of Physics ``A. Pontremoli", University of Milan, via Celoria 16, 20133 Milan, Italy}
\author{Alessio Zaccone}
\email{alessio.zaccone@unimi.it}
\affiliation[University of Milan]
{Department of Physics ``A. Pontremoli", University of Milan, via Celoria 16, 20133 Milan, Italy}

\author{Alberto Vailati}
\email{alberto.vailati@unimi.it}
\affiliation[University of Milan]
{Department of Physics ``A. Pontremoli", University of Milan, via Celoria 16, 20133 Milan, Italy}

\title[An \textsf{achemso} demo]
  {Convective Instability Driven by Diffusiophoresis of Colloids in Binary Liquid Mixtures}

\abbreviations{IR,NMR,UV}

\begin{tocentry}

\includegraphics[width=9cm,height=3.5cm,keepaspectratio]{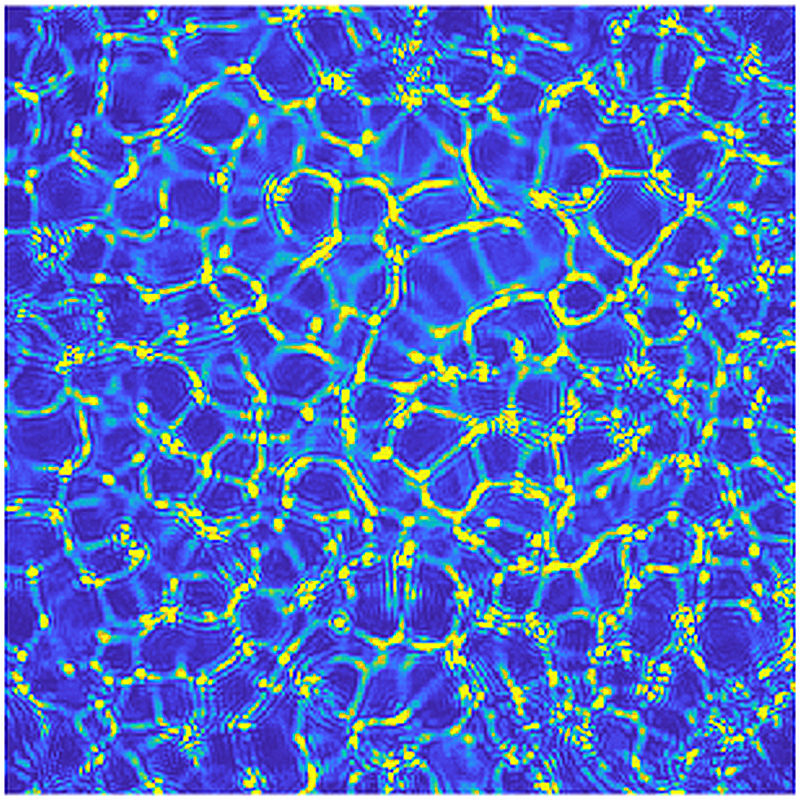}

\end{tocentry}

\begin{document}

%%%%%%%%%%%%%%%%%%%%%%%%%%%%%%%%%%%%%%%%%%%%%%%%%%%%%%%%%%%%%%%%%%%%%
%% The "tocentry" environment can be used to create an entry for the
%% graphical table of contents. It is given here as some journals
%% require that it is printed as part of the abstract page. It will
%% be automatically moved as appropriate.
%%%%%%%%%%%%%%%%%%%%%%%%%%%%%%%%%%%%%%%%%%%%%%%%%%%%%%%%%%%%%%%%%%%%%

%%%%%%%%%%%%%%%%%%%%%%%%%%%%%%%%%%%%%%%%%%%%%%%%%%%%%%%%%%%%%%%%%%%%%
%% The abstract environment will automatically gobble the contents
%% if an abstract is not used by the target journal.
%%%%%%%%%%%%%%%%%%%%%%%%%%%%%%%%%%%%%%%%%%%%%%%%%%%%%%%%%%%%%%%%%%%%%
\begin{abstract}
ABSTRACT: In a binary fluid mixture, the concentration gradient of a heavier molecular solute leads to a diffusive flux of solvent and solute to achieve thermodynamic equilibrium. If the solute concentration decreases with height, the system is always in a condition of stable mechanical equilibrium against gravity. We show experimentally that this mechanical equilibrium becomes unstable in case colloidal particles are dispersed uniformly within the mixture, and that the resulting colloidal suspension undergoes a transient convective instability with the onset of convection patterns. By means of a numerical analysis, we clarify the microscopic mechanism from which the observed destabilisation process originates. The solute concentration gradient drives an upward diffusiophoretic migration of colloids, in turn causing the development of a mechanically unstable layer within the sample, where the density of the suspension increases with height. Convective motions arise to minimize this localized rise in gravitational potential energy.  
\end{abstract}

%%%%%%%%%%%%%%%%%%%%%%%%%%%%%%%%%%%%%%%%%%%%%%%%%%%%%%%%%%%%%%%%%%%%%
%% Start the main part of the manuscript here.
%%%%%%%%%%%%%%%%%%%%%%%%%%%%%%%%%%%%%%%%%%%%%%%%%%%%%%%%%%%%%%%%%%%%%

A spontaneous drift motion of colloidal particles in a binary fluid mixture can be caused by the concentration gradient of a molecular solute \citep{derjaguin,DERJAGUIN_1993, Anderson1989,anderson_prieve}.  
This process, commonly named \textit{diffusiophoresis}, has found many applications in the last years, which include the  microfluidic separation of colloids \citep{palacci,Shin2020}, the colloidal self-assembly \citep{Shklyaev2018} and the motion of self-propelled nanoparticles \citep{Golestanian2005_PRL_microswimmers, Golestanian2007,Moran2017_phoreticselfpropulsion}.
Within this context, an increasing number of studies have investigated how to exploit diffusiophoresis for controlling the motion of dispersed particles in  porous media and dead-end channels \citep{Shin_PhysRevX2017, KAR_ACS_NANO_2015, SHIN_PNAS_2016, Akdeniz}, with particular emphasis to the quest of novel strategies for improving current particle manipulation capability \citep{BOLOGNESI_PhysRevLett.125.248002,Shi_PRL_2016}.
Very little attention has been paid, instead, to the implications of diffusiophoretic migration on the gravitational stability of colloidal suspensions. Yet this phenomenon appears to be crucial for particle manipulation in conditions where gravitational effects cannot be neglected, and to elucidate on outstanding problems such as the origin of the chaotic motion of catalytic particles \citep{chen_chong_liu_verzicco_lohse_2021} and of pattern formation in biological systems \citep{Alessio_Gupta_TURING,Ramm2021}.

\begin{figure}
\centering
\includegraphics[width = 0.9 \linewidth]{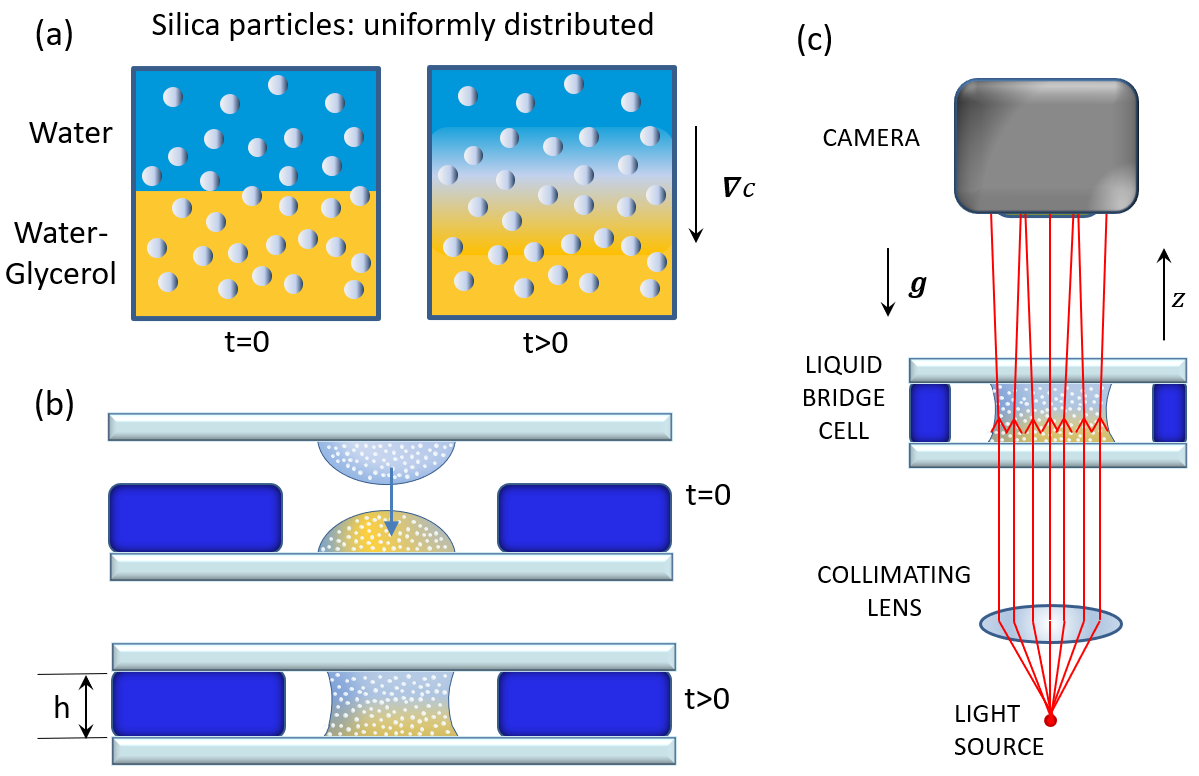}
\caption{(a) Conceptual scheme of the experiment: at time t=0  (left panel) a layer of water is brought into contact with a layer of a water-glycerol mixture. The two liquid phases contain silica particles with uniform volume fraction. As time goes by, the two liquid phases diffuse one into each other. (b) Liquid bridge experimental cell: at t=0 one drop of water and one of the water-glycerol mixture are deposited on two microscope glass slides and brought close to each other (top panel). The two droplets coalesce and form a liquid bridge (bottom panel);   (c) conceptual scheme of the shadowgraph optical setup: the light emitted by a LED is collimated by an achromatic doublet, crosses the liquid bridge cell in the vertical direction, and is eventually collected by a CMOS camera. } 
		\label{fig:setup}
\end{figure}

\begin{figure*}
\centering
\includegraphics[width = 0.9 \linewidth]{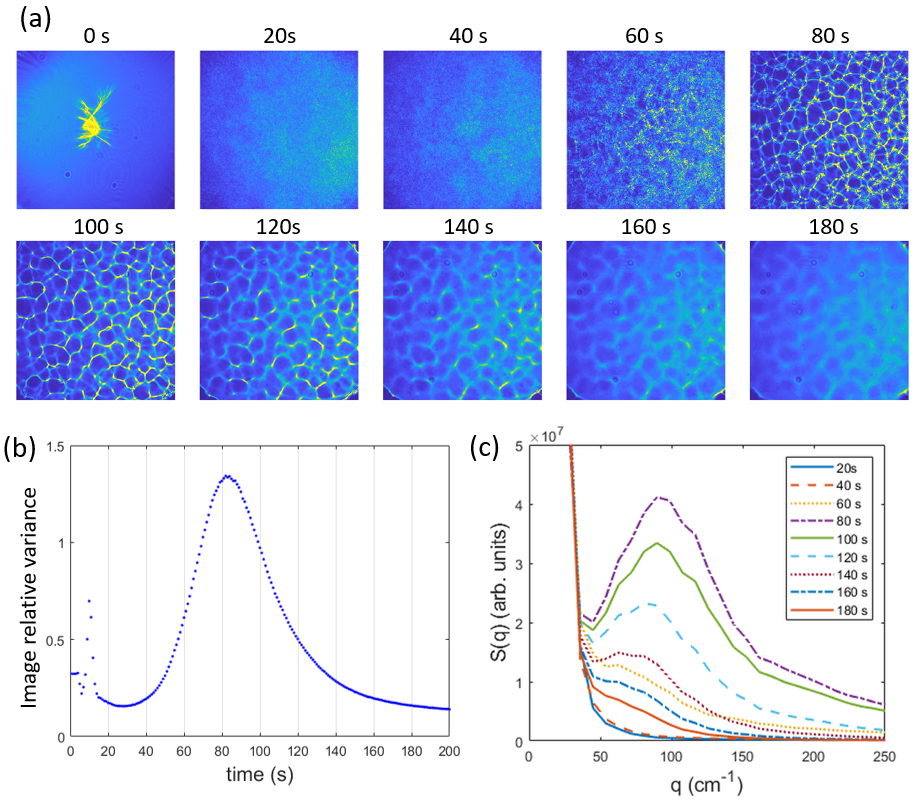}
	\caption{$\big($a - b$\big)$ Transient convective instability arising in a suspension of 22 nm diameter silica spheres uniformly dispersed (with colloid packing fraction $\phi_0=8\%$) within a stratified water-glycerol mixture. The glycerol displays a time-dependent, vertically decreasing, concentration profile (Fig. \ref{fig:setup}) and its mass fraction present on the bottom of the sample at $t=0$ is $c_0=56\%.$ (a) A sequence of shadowgraph images (side of each panel is 7.0 mm in real space) acquired in the interval of time $0$ s $\le t \le 180$ s shows the instability planform to consist of disordered cellular patterns which develop at $t_\textrm{s} \approx 60$ s, reach a maximum of contrast at  $t_\textrm{max} \approx 80$ s, and gradually fade away after 100 s. The relative variance $\sigma^2$, (b), and the power spectrum, (c), rapidly increase as a function of time until they are maximised at $t_\textrm{max} \approx 80$ s, and  decrease continuously for $t > t_\textrm{max}$.}
		\label{fig:comboseq}
\end{figure*}

In this Letter, we experimentally address the question of whether diffusiophoresis can lead a colloidal suspension towards a convective instability. We consider the case of silica nanospheres (LUDOX$\textregistered$ TMA, 22 nm diameter) uniformly dispersed in a fluid binary mixture in which a transient concentration gradient of molecular solute is produced by interdiffusion \citep{Oster1963}. More specifically, a layer of colloids in water is brought in contact from above with a layer of colloids dispersed at the same volume fraction in a water-glycerol mixture, and glycerol and water molecules are allowed to diffuse into each other to achieve thermodynamic equilibrium (Fig. \ref{fig:setup}(a)).
In the absence of dispersed colloids, the arranged system is always in a condition of stable mechanical equilibrium attained by layering the glycerol so that the overall density of the mixture decreases with height \citep{Castellini2023}. In this configuration, a small parcel of fluid displaced upwards by a perturbation becomes surrounded by fluid of smaller density, and this determines a restoring downward buoyancy force able to inhibit the onset of convective motions \citep{landau2013fluid, righetti}. When colloidal particles are distributed uniformly within the mixture, instead, we find that a convective instability arises in the system under generic conditions. This finding is intriguing because the onset of a convective instability requires the accumulation and minimization of gravitational potential energy, as occurs in the unstable configuration where the density of the fluid increases with height \citep{Velarde_RevModPhys}. In the suspension investigated here, instead, the potential energy is minimized due to the mechanically stable stratification of density determined by the glycerol, and no convection is expected to occur \citep{landau2013fluid,Veronis1968}. Since the concentration of the colloid is uniform, also the onset of double diffusion convection \citep{TREVELYAN2011} can be safely ruled out. 

To address this surprising observation, we numerically investigate the evolution of the base state of the system towards the instability and prove that a key role in the observed destabilisation process is played by the diffusiophoretic migration of colloids induced by the glycerol concentration gradient. We discuss the subtle mechanism through which diffusiophoresis is able to overturn the condition of stable mechanical equilibrium initially determined in the suspension by layering a heavier solute (glycerol) lying below a lighter solvent (water).
To perform experiments, we employed a custom cell recently developed by our group \citep{Giavazzi2016b}, which is based on the concept of a liquid bridge \citep{dG_capillary}, and allows one to bring  two miscible liquid phases into contact to start a diffusive process while simultaneously avoiding the formation of spurious perturbations at the liquid-liquid interface.
The cell is composed by two standard microscope glass slides placed one on top of the other (Fig. \ref{fig:setup}(b)). One of them is positioned horizontally on a stage anchored to an optical bench. A rubber spacer (Witte Blaue Matte) of thickness $h=0.95$ mm is placed on top of this slide to minimize sample evaporation during the experiments. The second slide is placed above the first one, at a distance larger than $h.$ A 37 $\mu$l droplet  of water-Ludox suspension (colloid volume fraction $\phi_0$) is anchored to the bottom of the upper slide, while another  droplet of water-glycerol-Ludox mixture (colloid volume fraction $\phi_0$ and glycerol mass fraction $c_0$) is deposited on the top of the lower slide. The upper slide is then shifted vertically by using a carriage so that the two drops are brought in contact and a liquid bridge forms between them. 
Patterns forming inside the sample are visualized using a shadowgraphy setup
 \citep{booksettles01} (Fig. \ref{fig:setup}(c)). A superluminous diode (SLD) of wavelength 675 $\pm$ 13 nm coupled to an optical fiber generates a diverging beam which is collimated on the sample by an achromatic doublet with focal length 10 cm. After crossing the cell, at a distance of 40 mm from it, the beam impinges on the sensor of a CMOS NX-S4 camera (IDT Vision)  with a resolution of 1024x1024 pixel and a 10-bit depth. 

A diffusive process starts as soon as the droplets come into contact. This instant 
represents the starting configuration of each measurement. 
Experiments are performed at room temperature under isothermal conditions, at fixed combinations ($\phi_0, c_0$), where $\phi_0$ is varied in the range $ \{1\% - 8 \%\} $ and $c_0$ in the range $ \{7 \% - 56 \% \}.$  For each measurement, a liquid bridge cell of the kind described above was prepared. Tests performed with tracer particles excluded the presence of Marangoni flow induced by the concentration gradient of the molecular solute at the free surface of the liquid bridge.
For each fixed concentration ($\phi_0, c_0$), starting from  $t=0,$ a sequence of 2000 images was acquired at a frequency of 1 Hz. A typical sequence of images, collected at $\phi_0= 8 \%$ and $c_0 = 56 \%,$ shows the development of an instability after a latency time of about $t_\textrm{s} \approx 60$ s (Fig. \ref{fig:comboseq}a)). The planform of the instability features a disordered cellular pattern, which reaches maximum contrast shortly after the onset at a time $t_\textrm{max} \approx$  80 s and gradually fades away in a time of the order of 100 s. A similar scenario is observed for all the other combinations of $\phi_0$ and $c_0$ (see Fig. S1 of the Supporting Information (SI) \citep{SI}).
In shadowgraph images, bright features are associated with regions where colloid concentration is higher (colloid rich regions act as positive focal length lenses) while dark features map colloid poor regions \citep{Cerbino2002_conv}. Hence, the strength of the convective patterns can be quantified through the 
 relative variance of an image defined as $\sigma^2 (t) \equiv \left\langle \big( I (\mathbf{x}, t) - \left\langle I (\mathbf{x}, t) \right\rangle \big)^2 \right\rangle/\left\langle I (\mathbf{x}, t) \right\rangle \big)^2,$ where $I (\mathbf{x}, t)$ is the intensity of light acquired by the pixel set at position $\mathbf{x} = (x,y)$ and the average $\left\langle \cdots \right\rangle$ is performed over all the pixels of the image. For each concentration $(\phi_0, c_0),$ the relative variance of the acquired sequence of shadowgraph images can be observed to rapidly increase as a function of time until a maximum $\sigma^2 (t_\textrm{max})$ is reached at $t_\textrm{max},$ and then to decrease for $t > t_\textrm{max}$ (Fig. \ref{fig:comboseq}(b)) and Fig. S2 of the SI \citep{SI}). An analogous behaviour as a function of time is shown, in correspondence to an unstable mode $ q_\textrm{peak},$ by the azimuthally averaged structure factor $S (q, t)$ associated to the acquired shadowgraph images (Fig. \ref{fig:comboseq}(c)) .

\begin{figure}
\centering
\includegraphics[width = 0.68 \linewidth]{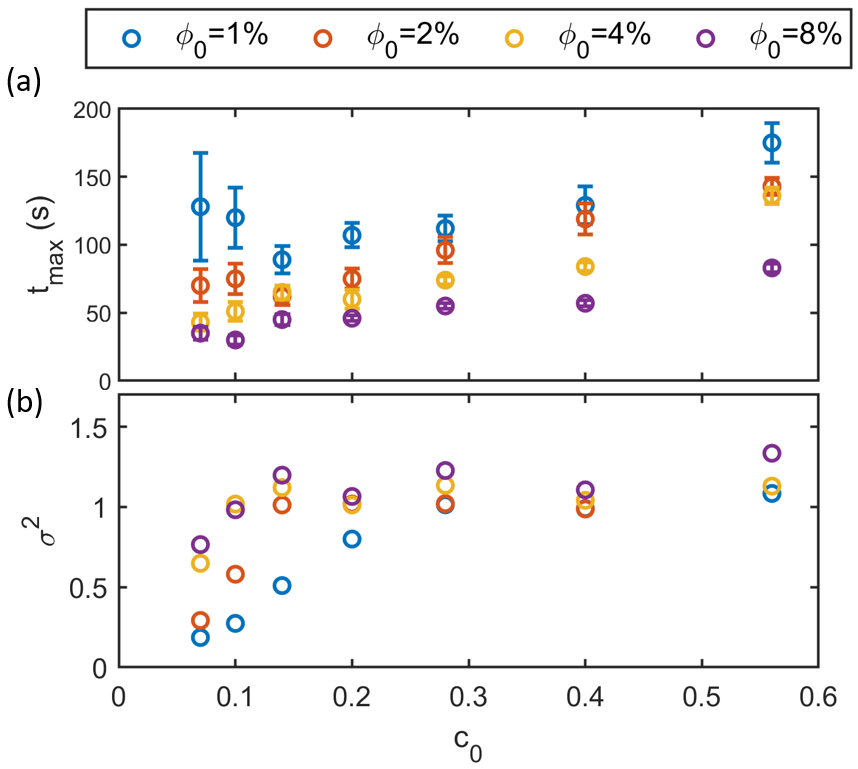}
\caption{ (a) experimentally measured time needed to reach peak variance $t_\textrm{max}$ and (b) relative variance  $\sigma^{2} (t_\textrm{max})$ of shadowgraph images, plotted as a function of $\phi_0$ and $c_0.$ When increasing $\phi_0$ at fixed values of $c_0,$ $t_\textrm{max}$ decreases while $\sigma^{2} (t_\textrm{max})$ increases. By contrast, at fixed values of $\phi_0,$ both $t_\textrm{max}$ and $\sigma^{2} (t_\textrm{max})$ increase as a function of $c_0.$}
\label{tmaxqmax}
\end{figure}

In order for the observed instability to arise, the system must be driven to a condition where, at least locally, its density increases with height. As temperature is kept constant in our experiments, the driving mechanism cannot be the fluid thermal expansion (such as in the Rayleigh-B\'enard convection \citep{Velarde_RevModPhys}) but rather an internal motion of colloids which, however, cannot be the free diffusion as at the beginning of each experiment the colloid concentration in our sample is uniform. Since an upward accumulation of colloidal particles in our system is only originated by the glycerol concentration gradient via diffusiophoresis, the latter must be the main responsible for the destabilization process.
In particular the proportionality, expected in nonelectrolyte systems \citep{Anderson_Lowell_Prieve_1982,Raj}, of the upward diffusiophoretic flux of colloids to both the colloid volume fraction and the strength of glycerol concentration gradient offers a simple explanation for the experimentally observed behaviour of $t_\textrm{max}$ and $\sigma^2 (t_\textrm{max})$ as a function of $\phi_0$ and $c_0$ (Fig. \ref{tmaxqmax}). $t_\textrm{max}$ decreases when increasing $\phi_0$ at fixed values of $c_0$ because a larger amount of nanoparticles accumulating towards the top of the sample results in the reduction of the time required for the appearance of a vertically increasing density profile in the suspension and, consequently, for the onset and the full development of the convection. To understand the increase of $t_\textrm{max}$ as a function of $c_0$ at fixed values of $\phi_0,$ instead, it must be considered that the increase of $c_0$ not only enhances the amount of migrating colloids due to diffusiophoresis, but also the density of the glycerol present on the bottom of the sample. A larger number of colloidal particles has to be transferred upwards to destabilize the system in this case, and the time required for the onset and the full development of the instability increases accordingly. The strength of the convective patterns, quantified by $\sigma^2 (t_\textrm{max}),$ increases as a function of both $\phi_0$ and $c_0,$ as in both cases the patterns involve the motion of a larger number of nanoparticles.

To support the hypothesis above, we performed a theoretical analysis of the evolution of the base state towards the instability.
Since the geometry of the experimental sample can be approximated with a cylinder of height $h=0.95$ mm and diameter $d \gg h,$ we introduce a 1D reference system with the $\mathbf{\hat{z}}$-axis pointing upward. We write the equation for the evolution of the base state as
\begin{equation} \label{base_state}
\begin{aligned}
&  \partial c (z, \tau)  / \partial \tau = ( D_1 / D_2 ) h^{2} \ \nabla^2 c (z, \tau), \\
&   \partial \phi (z, \tau)  / \partial \tau =  h^{2} \ \nabla \cdot \big( \nabla \phi (z, \tau)  + \alpha \phi (z, \tau) \nabla c (z, \tau) \big), 
\end{aligned}
\end{equation}
where $\nabla \equiv \partial / \partial z,$ while $c(z, \tau)$ and $\phi(z, \tau)$ indicate the spatial profiles of the glycerol mass fraction and of the colloid volume fraction, respectively, at dimensionless times $\tau \equiv D_2 t /h^2.$  $D_1$ and $D_2$ are the diffusion coefficients of glycerol in water and of colloidal spheres in a water-glycerol mixture, respectively. We treat the diffusiophoretic coefficient $\alpha$ as a ``free" parameter quantifying the strength of the diffusiophoretic coupling $\phi \nabla c.$ A meaningful value to $\alpha$ can be attributed by comparing theoretical and experimental results. 

\begin{figure}
\centering
\includegraphics[width = 0.95 \linewidth]{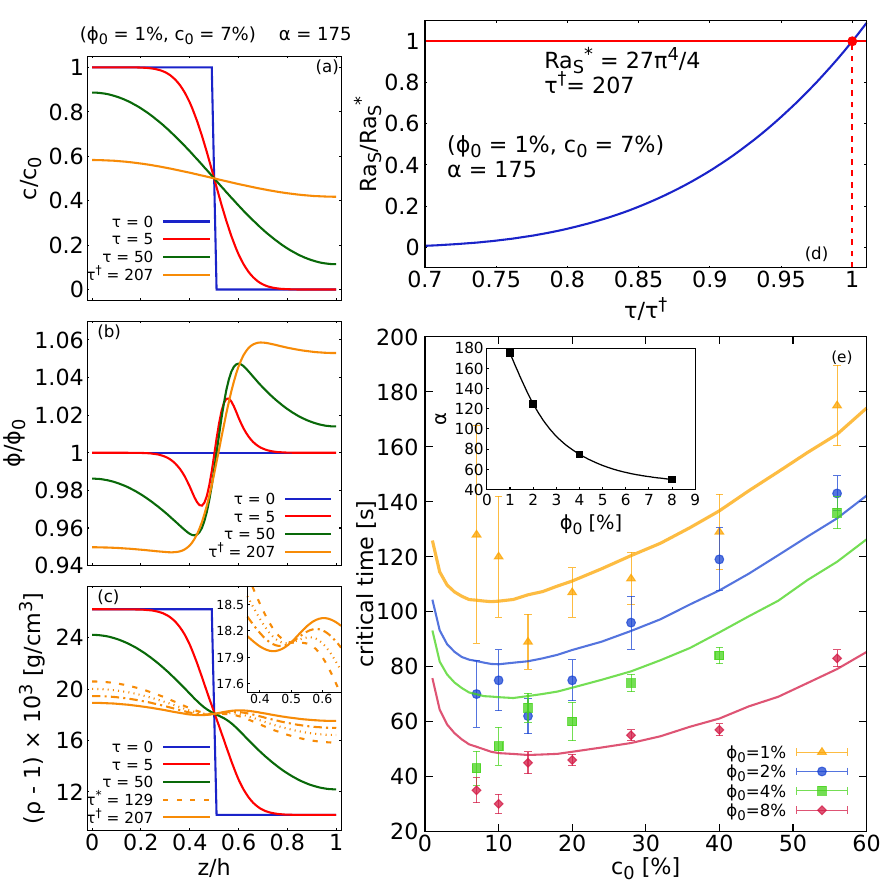}
\caption{Glycerol mass fraction $c$ (a), colloid volume fraction $\phi$ (b) and total density of the suspension $\rho$ (c) as a function of the $z$ coordinate of the sample in Fig. \ref{fig:setup}, for several dimensionless times $\tau \equiv D_2 t / h^2 \ge 0.$ Profiles are computed numerically at fixed ($\phi_0=1 \%, c_0= 7 \%$) and $\alpha=175.$ The height of the sample is $h=0.95$ mm and the value $D_2= 1.794 \times 10^{-7} \ $ cm$^2 \cdot$ s$^{-1}$ is assumed for the diffusion coefficient of colloids in the water-glycerol mixture.
The upward accumulation of colloids, driven by the concentration gradient of glycerol via diffusiophoresis, causes the slope of $\rho$ to display a local inversion after a time $\tau^* = 129$ (inset of (c)). (d) By associating a solutal Rayleigh number ($\textrm{Ra}_\textrm{s}$) to the density profiles at $\tau \ge \tau^*,$ an instability is detected at $\tau^{\dag}=207,$ when $\textrm{Ra}_\textrm{s}$ reaches the critical value $\textrm{Ra}_\textrm{s}^* = 27 \pi^4 /4.$ (e) For several combinations of $\phi_0$ and $c_0,$ after properly choosing $\alpha,$ this strategy leads to values of $\tau^\dag$ (lines)  in very good agreement with the experimental data of $t_\textrm{max}$ from Fig. \ref{tmaxqmax}(a) (points).}
\label{theory}
\end{figure}

We apply the Neumann (zero-flux) boundary conditions $(\partial c / \partial z)_{z=0} = (\partial c / \partial z)_{z=h} =0$ and $(\partial \phi / \partial z)_{z=0} = (\partial \phi / \partial z)_{z=h} =0.$ As initial conditions, we assume  $c(z, \tau=0)= c_0$ for $0 \le z/h \le 0.5$ and $c(z, \tau=0)= 0$ for $0.5 < z/h \le 1,$ while we consider $\phi(z, \tau =0) = \phi_0 $ for $0 \le z/h \le 1.$ 
We hence solve the system \eqref{base_state} for many combinations of $\alpha,$ $c_0$ and $\phi_0$ \bibnote{At any time $\tau,$ the top equation of Eqs. \eqref{base_state} is solved analytically (via separation of variables \citep{crank}). After insertion of the obtained $c(z, \tau),$ the bottom equation of Eqs. \eqref{base_state} is solved numerically (via finite difference method \citep{crank,smith1985numerical}) to compute $\phi (z, \tau)$}. We use data from  Ref. \citep{nishijima} to fix $D_1$ and the Stokes-Einstein formula to compute $D_2$ \bibnote{We use data from  Ref. \citep{nishijima} to fix $D_1$ and the Stokes-Einstein formula to compute $D_2 = k_B T /(6 \pi \eta a)$, where $k_B$ is the Boltzmann constant, $a= 11$ nm the radius of the colloidal particles, and for the shear viscosity $\eta$ of the water-glycerol mixture we use data collected at  $T=20^{\circ}$C \citep{lide2004crc}. $D_1$ and $D_2$  vary individually as a function of $c_0,$ however their ratio remains constant to $D_1 /D_2  \approx 100$ for any given glycerol mass fraction. The coefficients $\rho_0$ and $\beta$ in the expression to compute the density profile of the suspension are obtained through a fit of data from Ref. \citep{lide2004crc}. Slightly different values are obtained by varying $c_0,$ which are listed in Table S1 of the SI \citep{SI}. The values of $D_1,$ $D_2,$ $\rho_0$ and $\beta$ never depend on $\phi_0$}. The same values considered in experiments are chosen for $c_0$ and $\phi_0.$
The main goal of our theoretical analysis is to understand whether the diffusiophoretic term $\alpha \phi \nabla c$ in Eqs. \eqref{base_state} is able to drive the base state towards a mechanically unstable configuration, a task which requires having access to the time evolution of the density profile of the suspension. 
The latter, for any solution $c(z, \tau)$ and $\phi(z, \tau)$ of system \eqref{base_state}, is given by (see SI \citep{SI}) $\rho (c,\phi) = \rho_0 + (\rho_L - \rho_0) \phi + \beta \big( 1 - \phi \big) c,$ where $\rho_0$ is the density of the solvent in the absence of glycerol molecules, $\rho_L = 2.2 \ \textrm{g}/\textrm{cm}^3$ is the density of a single colloidal particle, and $\beta \equiv d \rho_{B} /d c$ is the solutal expansion coefficient of a water-glycerol mixture of density $\rho_B$. 
For a combination $(\phi_0, c_0)$ at fixed $\alpha \neq 0,$ glycerol starts diffusing upwards when $\tau >0$ (Fig. \ref{theory}a)). Simultaneously, colloids gradually deplete from the bottom half of the cell and accumulate towards the top half (Fig. \ref{theory}(b)). This colloid migration, initially located in the central region
of the sample where the concentration gradient of the glycerol is the largest, has relevant repercussions on the overall density profile of the suspension. After a time $\tau^*,$ indeed, $\rho$ develops a local minimum  $\rho_\textrm{min}$ at a height $z_\textrm{min},$ followed by a local maximum $\rho_\textrm{max}$ at a height $z_\textrm{max}>z_\textrm{min}$ (Fig. \ref{theory}(c)).
At any $\tau \ge \tau^*,$ gravity acts as to pull denser fluid parcels between $z_\textrm{min}$ and $z_\textrm{max}$ from the top to the bottom, and is opposed by the viscous damping force of the fluid. Following Ref. \citep{CALTECH}, we quantify the competition between these two opposite effects by means of a \textit{solutal} Rayleigh number $\textrm{Ra}_\textrm{s} \equiv g (\Delta z)^3 \Delta \rho / (\eta D_2),$ where $\Delta z \equiv z_\textrm{max} - z_\textrm{min},$ $\Delta \rho \equiv \rho_\textrm{max} - \rho_\textrm{min},$ $g$ is the gravity acceleration, $\eta$ is the viscosity of the water-glycerol mixture and (again) we compute $D_2$ via the Stokes-Einsten formula. The increase of $\textrm{Ra}_\textrm{s}$ with time (Fig. \ref{theory}(d)), which follows from that of $\Delta z$ and $\Delta \rho,$ signals the growing importance of the destabilizing effect of gravity in our system. The latter is hence driven towards a convective instability.
For several combinations of $\phi_0$ and $c_0,$ we can obtain an estimate of the ``critical'' time $\tau^\dag$ at which the instability sets in by using the threshold value  $\textrm{Ra}_\textrm{s}^{*} = 27 \pi^4 /4$ associated to isothermal and free boundary conditions \citep{Messlinger2013}.
When $\alpha$ is properly chosen, a good agreement between  theoretical curves for $\tau^\dag (c_0) h^2/ D_2$ and experimental data for $t_\textrm{max} (c_0)$ is found for each $\phi_0$ (Fig. \ref{theory}(e)). For comparison we rely on $t_\textrm{max}$ rather than on the time $t_\textrm{s}$ at which patterns start forming, since the former instant is typically more clearly identifiable in a plot of the time evolution of the relative variance (Fig. \ref{fig:comboseq}(b)).
A general feature of  $\tau^\dag$ when plotted as a function of $c_0$ at fixed values of $\phi_0$ and $\alpha$, is that it exhibits a minimum at $c_0\approx 7\%$ (Fig. \ref{theory}(e)), which thus represents  the characteristic glycerol concentration that maximizes the efficiency of the diffusiophoretic process in competition to the stabilizing solvent profile and leads to a minimization of the time needed for the onset of the instability.

Overall, fulfillment of the condition $\tau^\dag h^2 / D_2 \approx  t_\textrm{max}$ at fixed concentrations $(\phi_0, c_0)$ can be regarded as a procedure to attribute a meaningful value to the diffusiophoretic mobility coefficient $\alpha.$ In other words, for a given curve $\tau^\dag (c_0)$ at a fixed $\phi_0$, the value of $\alpha$ can be extracted by requiring that $\tau^\dag$ (expressed in proper units) is compatible with the  time $t_\textrm{max}$ experimentally measured for that combination. Since inter-colloids interaction are not taken into account in Eqs. \eqref{base_state}, this procedure is only strictly rigorous in the limit $\phi \to 0$ where a ``bare" value $\alpha_b$ is obtained. In the most diluted among the experimentally studied cases, $\phi_0 = 1 \%,$ we find $\alpha_b \approx 175,$ a best-fitting value on the same order of magnitude of $D_1/D_2\approx 100.$ Remarkly, this finding is compatible with recent observations reported within the context of colloid diffusiophoresis driven by the concentration gradient of an electrolyte \citep{Gupta2020}, where theoretical modelling and microfluidics experiments have shown the ambipolar diffusivity of the solvent to represent an upper bound for the diffusiophoretic mobility. Our results suggest that a similar relation could apply also to diffusiophoresis in nonelectrolitic systems, without any signature of the super-diffusion or arrested spreading effects predicted in Ref. \citep{Chu2020}.

For $\phi_0 > 1 \%,$ we observe a good agreement between the experimental curves for $\tau_\textrm{max} (c_0)$ and the theoretical curves for $\tau^\dag (c_0)$ in correspondence of values of $\alpha$ smaller than $\alpha_b$ (inset of Fig. \ref{theory}(e)). More in detail, an increase in $\phi_0$ of a factor 8 requires to reduce $\alpha$ by a factor 3-4 to attain a meaningful agreement with the experimental results. We attribute the observed behavior of the diffusiophoretic mobility coefficient to the neglection of inter-colloids interactions in Eqs. \eqref{base_state}, an approximation only reasonable in diluted conditions where $\phi_0 \to 0.$ At larger values of $\phi_0,$ instead, when combined with the mapping procedure described above, this simplifying assumption leads to effective values of $\alpha$ that embed the contribution of the inter-colloids interactions. It seems that the latters, which are always present in experiments, are able to reduce the diffusiophoretic coupling $\alpha \phi \nabla c.$ This reduction, however, does not affect the decrease of $\tau^\dag$ when $\phi_0$ is increased for fixed values of $c_0$ (Fig. \ref{theory}(e)).

In summary, we demonstrate the existence of a novel convective instability occurring under isothermal conditions in a suspension of colloidal particles dispersed uniformly within a fluid mixture where the concentration of a molecular solute decreases with height. This result is surprising because the suspension is originally in a condition of stable mechanical equilibrium determined by having a heavier solute lying below a lighter solvent, and convective flows are consequently inhibited. The key point to understand this striking outcome is the dominance, for single colloidal particles, of surface tension effects over gravitational and Brownian forces. To reduce the solute-colloid interfacial energy \citep{RUCKENSTEIN}, indeed, particles experience an upward diffusiophoretic motion which leads them to deplete from the bottom of the sample and to accumulate toward the top. Consequently, a mechanically unstable layer develops within the sample where the density of the suspension increases with height and a convective instability arises to minimize this localized rise in gravitational potential energy.
Our results have significant implications for the control of pattern formation and for the handling of nanoparticles in inhomogeneous environments, and thus widespread applications in industrial, geological and biological systems. Understanding how the reported scenario varies under  reduced gravity and hyper gravity conditions  represents a future ambitious task of strategic relevance for space exploration \citep{Vailati2023,Vailati2024}.

%\section{Biographies}

%%%%%%%%%%%%%%%%%%%%%%%%%%%%%%%%%%%%%%%%%%%%%%%%%%%%%%%%%%%%%%%%%%%%%
%% The "Acknowledgement" section can be given in all manuscript
%% classes.  This should be given within the "acknowledgement"
%% environment, which will make the correct section or running title.
%%%%%%%%%%%%%%%%%%%%%%%%%%%%%%%%%%%%%%%%%%%%%%%%%%%%%%%%%%%%%%%%%%%%%
\begin{acknowledgement}

Work partially supported by the European Space Agency (ESA) in the framework of the ``Giant fluctuations'', NESTEX and ``Sedimenting Colloids'' projects, and by the Italian Space Agency (ASI) through the projects ``Gravitationally TApping Colloids in Space (GTACS) - Sedimenting Colloids'' (Number 2023-19-U.0) and ``Non-Equilibrium Phenomena in Soft Matter and Complex Fluids (NESTEX)'' (Number 2023-20-U.0).
A.Z. gratefully acknowledges funding from the European Union through Horizon Europe ERC Grant number: 101043968 ``Multimech'', from US Army Research Office through contract nr.   W911NF-22-2-0256, and from the Nieders{\"a}chsische Akademie der Wissenschaften zu G{\"o}ttingen in the frame of the Gauss Professorship program. 

\end{acknowledgement}

%\suppinfo{Images of convective patterns at peak contrast and time evolution of the contrast of the convective patterns  for all the explored concentration combinations; details about the theoretical model (PDF).}

\appendix
\section{Supplementary Information}
\section{Convective patterns}

\begin{figure*}
   \centering
   \includegraphics[width=15cm]{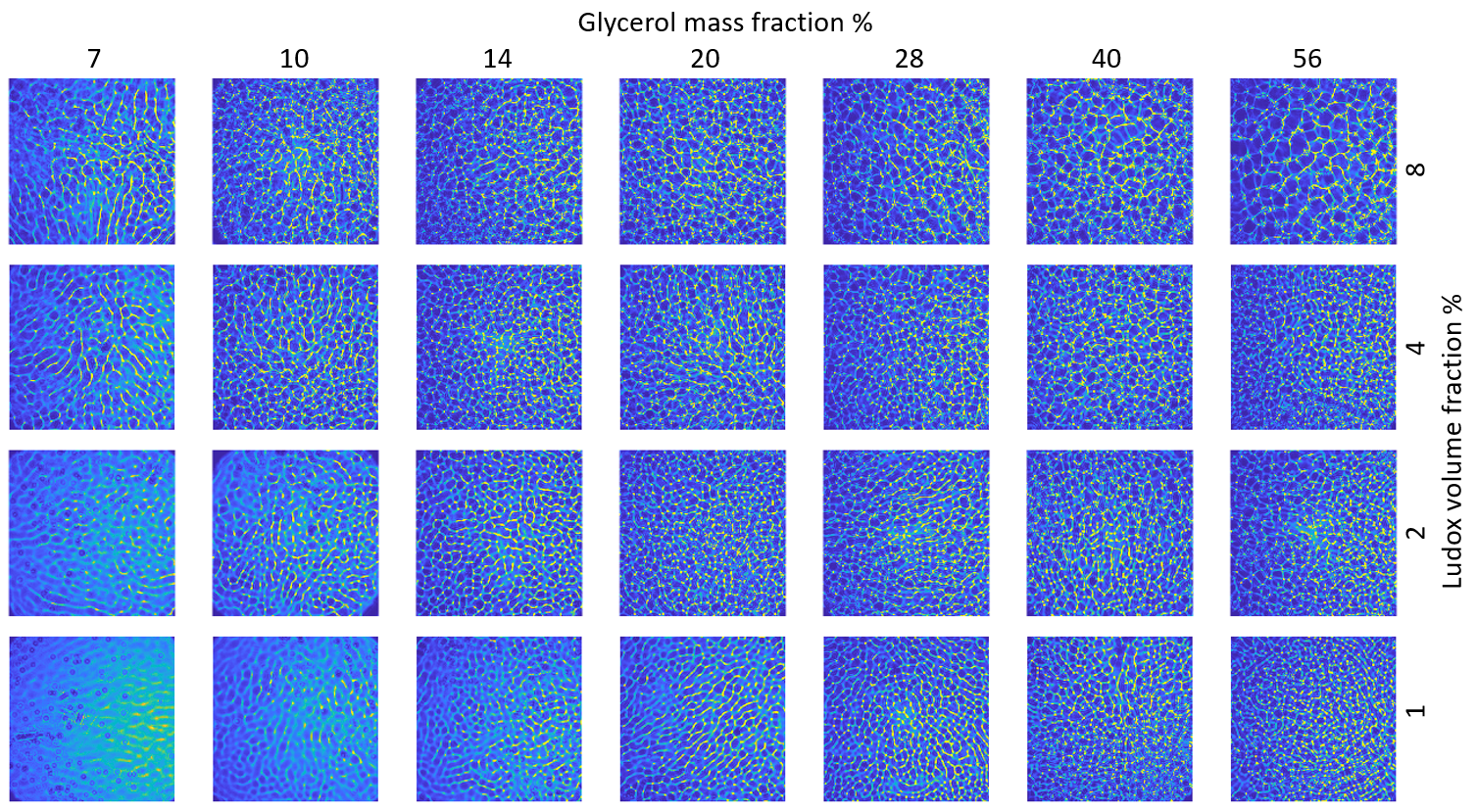}
   \caption{Shadowgraph images corresponding to the $t_\textrm{max}$ of Fig. \ref{variance_SI}. Each image shows the planform of the instability, at its maximum strength of intensity, to consist of a disordered cellular pattern. The panel of each image corresponds to 7.0 mm in real space.}
   \label{pattern_SI}
\end{figure*}

\begin{figure*}
   \centering
   \includegraphics[width=15cm]{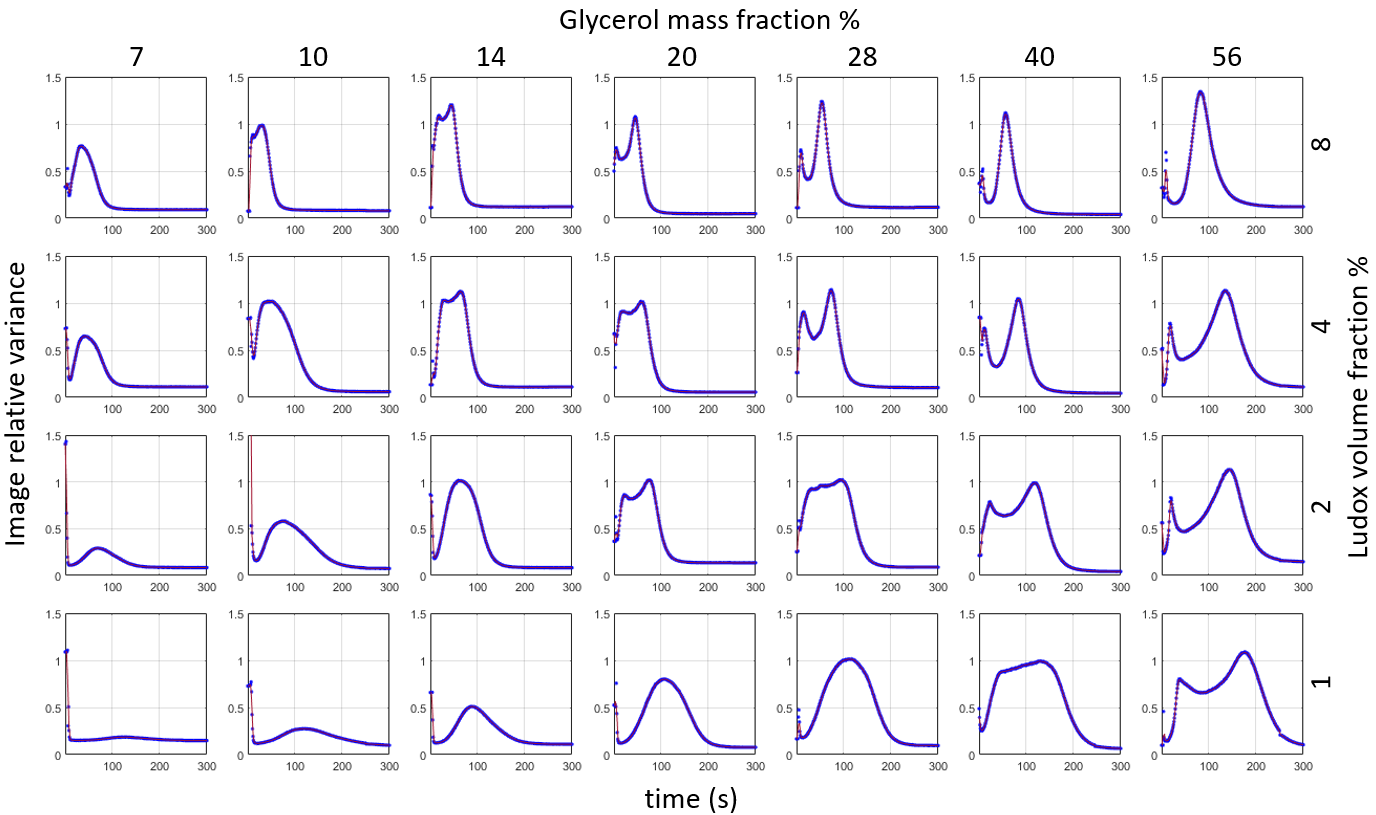}
   \caption{Temporal evolution of the relative variance for several values of the Ludox volume fraction and the glycerol mass fraction at time $t=0$ (see main Letter). Each image shows the relative variance to increase until a maximum is reached at a time $t_\textrm{max}.$}
   \label{variance_SI}
\end{figure*}

Pattern formation has been investigated systematically for  combinations of the colloid volume fraction  $\phi_0$ in the range $ \{1\% - 8 \%\} $ and of the concentration of glycerol $c_0$ in the range $ \{7 \% - 56 \% \}$. In all the investigated cases convective patterns develop after a latency time. Figure \ref{pattern_SI} shows the patterns when they have reached maximum contrast. The time evolution of the contrast of the convective patterns shows the emergence of a  separation of the colloid, followed by a mixing phase determined by the onset of transient convective motions (Fig. \ref{variance_SI}).

\clearpage

\section{Theoretical model for the evolution of the base state}

\subsection{Diffusion coefficients}

To solve Eqs. (1) of the main Letter, the ratio $D_1 / D_2$ between the diffusion coefficients of glycerol in water, $D_1,$ and of colloidal spheres in a water-glycerol mixture, $D_2,$ has to be specified. In general, independently on the amount of dispersed colloids, these two coefficients are expected to vary as a function of the mass fraction of glycerol $c_0$ present in the bottom layer at $t=0$ (see Fig. 1 of the main Letter). We recall that experimental values of this latter quantity are used as initial conditions in our theoretical model. For any given value of $c_0,$ the glycerol mass fraction $c$ varies along the (vertical) $z$ direction (see Fig. 4(a) of the main Letter) and, consequently, multiple values of $c$ exist which could be exploited to fix the values of $D_1$ and $D_2$ corresponding to that $c_0.$ To deal with this issue, we adopt the following strategy: for any given value of $c_0,$ we compute $D_1$ and $D_2$ at $\bar{c}=c_0/2.$ In other words, we fix $D_1 \equiv D_1 (\bar{c})$ and $D_2 \equiv D_2 (\bar{c}).$ 

We extrapolate values of $D_1$ from Ref. \citep{nishijima}. Instead, we compute $D_2$ by means of the Stokes-Einstein formula 
\begin{equation}
\label{stokes_einstein_SM}
D_2 = \frac{k_B T}{6\pi \eta a}.
\end{equation}
Here $k_B$ is the Boltzmann constant, $T$ is the temperature, $a$ is the radius of the diffusing colloids and $\eta$ is the viscosity of the water-glycerol mixture. From Eq. \eqref{stokes_einstein_SM}, it is clear that the dependence of $D_2$ on $c_0$ is embedded in the dependence of $\eta$ on $c_0.$ Again, for any $c_0,$ the corresponding value of the viscosity is fixed to $\eta \equiv \eta (\bar{c})$ where $\bar{c}= c_0/2.$ We extract $\eta$ from data concerning a water-glycerol mixture at $T= 20^{\circ}$ reported in Ref. \cite{lide2004crc}. To compute $D_2$ we recall that $a= 11$ nm in our experimental system.

For all the values of $c_0$ considered in our experiments, while $D_1$ and $D_2$ individually vary as a function of $c_0,$ their ratio remains constant $D_1/D_2 \approx 100.$ This is the value which is used to solve Eqs. (1) of the main Letter. %However the data of $D_2$ listed in the last column of Table are important for the conversion between times $\tau \equiv $

\subsection{Density profile}

The mass density $\rho$ of our system is a function of the position $\mathbf{r}$ and of the time $t.$ i. e. $\rho \equiv \rho (\mathbf{r},t).$ At a fixed time $t,$ the total mass of fluid present in an elementary volume element $V$ centered at $\mathbf{r}$ is given by $M= M_L + M_B,$ where $M_L$ and $M_B=M_\textrm{water} + M_\textrm{glycerol}$ are the masses of the Ludox particles and of the binary water-glycerol mixture contained in $V,$ respectively. It follows that
\begin{equation} \label{density_SM}
\rho \equiv \frac{M}{V} = \frac{M_L}{V} + \frac{M_B}{V} .
\end{equation}
Calling $N_L$ the number of Ludox particles contained in $V,$ and $m_L$ their mass, it follows that $M_L=N_L m_L.$ Moreover, calling $v_L$ the volume of each colloid, the volume $V_L$ occupied by the Ludox particles within $V$ can be written as $V_L = N_L v_L.$ It follows that the term $M_L/V$ in Eq. \eqref{density_SM} can be written as
\begin{equation} \label{rhs_1}
\frac{M_L}{V} = \phi \rho_L, 
\end{equation}
where $\phi \equiv V_L/V$ is the colloid volume fraction and $\rho_L \equiv m_L/v_L$ is the density of a single colloidal particle.

To determine the term $M_B/V$ in Eq. \eqref{density_SM}, we notice that the part of $V$ available for the water molecules and the glycerol particles is $V_B=V-V_L.$ It follows that the density of the water-glycerol mixture is $\rho_B = M/(V-V_L),$ and that
\begin{equation} \label{rhs_2}
\frac{M_B}{V} = \rho_B \frac{V-V_L}{V} = \rho_B (1-\phi).
\end{equation}
Inserting Eqs. \eqref{rhs_1} and \eqref{rhs_2} into Eq. \eqref{density_SM} we obtain
\begin{equation} \label{insiem}
\rho = (\rho_L-\rho_B) \phi + \rho_B.
\end{equation}
Since $\rho_B$ is a function of $c,$ the dependence of $\rho$ on $c$ in Eq. \eqref{insiem} is contained in $\rho_B.$
Experimental data from Ref. \citep{lide2004crc} show that 
 $\rho_B$ exhibits a nearly linear dependence on $c$: 
\begin{equation} \label{SM_penultima}
    \rho_B = \rho_0 + \beta c,
\end{equation}
where $\rho_0$ is the density of the solvent in the absence of glycerol molecules, and $\beta \equiv d \rho_B / d c$ is the solutal expansion coefficient of a water-glycerol mixture of density $\rho_B.$ 

Inserting Eq. \eqref{SM_penultima} into Eq. \eqref{insiem}, we get
\begin{equation} \label{SM_ultima}
    \rho (c, \phi) = \rho_0 + (\rho_L - \rho_0) \phi + \beta (1-\phi) c.
\end{equation}

To take into account small deviations from linearity of the experimental data of Ref. \citep{lide2004crc}, in Eq. S\ref{SM_penultima} we assume that $\beta$ depends on $c_0$. We characterize this dependence
 through a fit of the data from Ref. \citep{lide2004crc} in the  concentration range $[0, c_0]$. Slightly different values of $\beta$ are obtained by varying $c_0,$ which are listed in Table \ref{table1}. To check the consistency of this method, during the fitting procedure we also leave $\rho_0$ as a free parameter and, as expected, we obtain a nearly constant value of $\rho_0$, corresponding to the density of pure water (Table \ref{table1}). The values of $\rho_0$ and $\beta$ are assumed to be independent on $\phi_0.$

\begin{table}[b]
\centering % used for centering table
\begin{tabular}{c c c c c c c c} % centered columns (4 columns)
\hline\hline %inserts double horizontal lines
$c_0$ [\%] & \ $\rho_0$ [g\, cm$^{-3}$] & \ $\beta$ [g\, cm$^{-3}$]  \\ [0.5ex] % inserts table
%heading
\hline % inserts single horizontal line
  7 & 0.9982 & 0.2305 \\ % inserting body of the table
 10 & 0.9982 & 0.2332 \\ % inserting body of the table
 14 & 0.9981 & 0.2340 \\ % inserting body of the table
 20 & 0.9979 & 0.2376 \\ % inserting body of the table
 28 & 0.9977 & 0.2418 \\ % inserting body of the table
 40 & 0.9972 & 0.2487 \\ % inserting body of the table
 56 & 0.9965 & 0.2554  \\ % [1ex] % [1ex] adds vertical space
\hline \hline
\end{tabular}
\caption{Values of $\rho_0$ and $\beta$ in Eq. \eqref{SM_ultima} as a function of $c_0.$ The values are obtained through a fit to data from Ref.  \citep{lide2004crc} in the concentration range $[0,c_0]$.} 
\label{table1} % is used to refer this table in the text
\end{table}

%%%%%%%%%%%%%%%%%%%%%%%%%%%%%%%%%%%%%%%%%%%%%%%%%%%%%%%%%%%%%%%%%%%%%
%% The same is true for Supporting Information, which should use the
%% suppinfo environment.
%%%%%%%%%%%%%%%%%%%%%%%%%%%%%%%%%%%%%%%%%%%%%%%%%%%%%%%%%%%%%%%%%%%%%

%%%%%%%%%%%%%%%%%%%%%%%%%%%%%%%%%%%%%%%%%%%%%%%%%%%%%%%%%%%%%%%%%%%%%
%% The appropriate \bibliography command should be placed here.
%% Notice that the class file automatically sets \bibliographystyle
%% and also names the section correctly.
%%%%%%%%%%%%%%%%%%%%%%%%%%%%%%%%%%%%%%%%%%%%%%%%%%%%%%%%%%%%%%%%%%%%%
\bibliography{bib_JPCL_anzivino_rev1}

\end{document}